\documentstyle[psfig,letters,referee]{mn}
\newif\ifAMStwofonts
\def\ang{{$\AA$}\ }
\def\dfe{{$\frac{df}{d\epsilon}$}\ }
\def\far{{${f_r}$}\ }
\def\chan{{\it Chandra}\ }
\def\xmm{{\it XMM-Newton}\ }
\def\mcg{{MCG$-$6-30-15}\ }

\def\efe25{{$(e~+~Fe~XXV) \rightarrow Fe~XXIV$}\ }
\def\o1{{O \rm I\ }}
\def\o2{{O \rm II\ }}
\def\o3{{O \rm III\ }}
\def\o4{{O \rm IV\ }}
\def\o5{{O \rm V\ }}
\def\o6{{O \rm VI\ }}
\def\o7{{O \rm VII\ }}
\def\o8{{O \rm VIII\ }}
\def\ka{{K$\alpha$\ }}
\def\etal{{\it et\thinspace al.}\ }

\def\lmlm{{$\lambda\lambda$}\ }

\def\eion{{(e~+~ion)}\ }
\newcommand{\be}{\begin{equation}}
\newcommand{\ee}{\end{equation}}
\newcommand{\ba}{\begin{array}{l}}
\newcommand{\ea}{\end{array}}
\newcommand{\lra}{{\longrightarrow}}
\ifoldfss
  \ifCUPmtlplainloaded \else
    \NewTextAlphabet{textbfit} {cmbxti10} {}
    \NewTextAlphabet{textbfss} {cmssbx10} {}
    \NewMathAlphabet{mathbfit} {cmbxti10} {} % for math mode
    \NewMathAlphabet{mathbfss} {cmssbx10} {} %  "   "    "
  \fi
  \ifAMStwofonts
    \ifCUPmtlplainloaded \else
      \NewSymbolFont{upmath} {eurm10}
      \NewSymbolFont{AMSa} {msam10}
      \NewMathSymbol{\upi}     {0}{upmath}{19}
      \NewMathSymbol{\umu}     {0}{upmath}{16}
      \NewMathSymbol{\upartial}{0}{upmath}{40}
      \NewMathSymbol{\leqslant}{3}{AMSa}{36}
      \NewMathSymbol{\geqslant}{3}{AMSa}{3E}

    \fi
  \fi
\fi % End of OFSS

\ifnfssone
  \newmathalphabet{\mathit}
  \addtoversion{normal}{\mathit}{cmr}{m}{it}
  \addtoversion{bold}{\mathit}{cmr}{bx}{it}
  \newmathalphabet{\mathbfit} % math mode version of \textbfit{..}
  \addtoversion{normal}{\mathbfit}{cmr}{bx}{it}
  \addtoversion{bold}{\mathbfit}{cmr}{bx}{it}
  \newmathalphabet{\mathbfss} % math mode version of \textbfss{..}
  \addtoversion{normal}{\mathbfss}{cmss}{bx}{n}
  \addtoversion{bold}{\mathbfss}{cmss}{bx}{n}
  \ifAMStwofonts
    \ifCUPmtlplainloaded \else
      \UseAMStwoboldmath
      \makeatletter
      \new@mathgroup\upmath@group
      \define@mathgroup\mv@normal\upmath@group{eur}{m}{n}
      \define@mathgroup\mv@bold\upmath@group{eur}{b}{n}
      \edef\UPM{\hexnumber\upmath@group}
      \new@mathgroup\amsa@group
      \define@mathgroup\mv@normal\amsa@group{msa}{m}{n}
      \define@mathgroup\mv@bold\amsa@group{msa}{m}{n}
      \edef\AMSa{\hexnumber\amsa@group}
      \makeatother
      \mathchardef\upi="0\UPM19
      \mathchardef\umu="0\UPM16
      \mathchardef\upartial="0\UPM40
      \mathchardef\leqslant="3\AMSa36
      \mathchardef\geqslant="3\AMSa3E
    \fi
  \fi
\fi % End of NFSS release 1

\ifnfsstwo
  \DeclareMathAlphabet{\mathbfit}{OT1}{cmr}{bx}{it}
  \SetMathAlphabet\mathbfit{bold}{OT1}{cmr}{bx}{it}
  \DeclareMathAlphabet{\mathbfss}{OT1}{cmss}{bx}{n}
  \SetMathAlphabet\mathbfss{bold}{OT1}{cmss}{bx}{n}
  \ifAMStwofonts
    \ifCUPmtlplainloaded \else
      \DeclareSymbolFont{UPM}{U}{eur}{m}{n}
      \SetSymbolFont{UPM}{bold}{U}{eur}{b}{n}
      \DeclareSymbolFont{AMSa}{U}{msa}{m}{n}
      \DeclareMathSymbol{\upi}{0}{UPM}{"19}
      \DeclareMathSymbol{\umu}{0}{UPM}{"16}
      \DeclareMathSymbol{\upartial}{0}{UPM}{"40}
      \DeclareMathSymbol{\leqslant}{3}{AMSa}{"36}
      \DeclareMathSymbol{\geqslant}{3}{AMSa}{"3E}
    \fi
  \fi
\fi % End of NFSS release 2

\ifCUPmtlplainloaded \else
  \ifAMStwofonts \else % If no AMS fonts
    \def\upi{\pi}
    \def\umu{\mu}
    \def\upartial{\partial}
  \fi
\fi

\title{X-ray absorption via K$\alpha$ resonance complexes in oxygen ions}
\author[Pradhan \etal]
       {Anil K. Pradhan$^1$,Guo Xin Chen$^1$,Franck Delahaye$^1$,Sultana
N. Nahar$^1$ \& Justin Oelgoetz$^2$\\
       $^1$ Department of Astronomy, $^2$ Department of Chemistry,
 The Ohio State University, Columbus, OH 43210, USA}
\date{Accepted  xxxxxx 
      Received xxxxxx;
      in original form xxxxxx}

\pagerange{\pageref{firstpage}--\pageref{lastpage}}
\pubyear{1994}

\def\LaTeX{L\kern-.36em\raise.3ex\hbox{a}\kern-.15em
    T\kern-.1667em\lower.7ex\hbox{E}\kern-.125emX}

\begin{document}

\maketitle

\label{firstpage}
\begin{abstract}
 The \ka resonance complexes in oxygen ions O~I - O~VI are theoretically
computed and resonance oscillator strengths and wavelengths are presented. 
The highly resolved photoionization cross sections, with relativistic
fine structure, are  computed
in the coupled channel approximation using the Breit-Pauli R-matrix method.
A number of strong \ka resonances are found to
be appreciable, with resonance oscillator strengths $f_r > 0.1$. 
The \ka resonance wavelengths of O~I-O-VI 
lie in a relatively narrow wavelength range $22 - 23.5 \AA$, and
the X-ray opacity in this region should therefore be significantly affected
by K $\rightarrow$ L transitions in oxygen. The results should be
useful in the interpretation of soft X-ray spectra observed from \chan and \xmm.
\end{abstract}
\section{Introduction}
 K-shell excitation leads to the formation of resonance complexes in all 
atomic systems, with the exception of H- and He-like ions. The \ka
complex in particular, associated with the 1s $\lra$ 2p transition, is
expected to be generally strong. The same transitions in H- and
He-like ions are well known and have been observed in many \chan and
\xmm sources (e.g. Branduardi-Raymont \etal 2001, Lee \etal 2001, 
Ness \etal 2001, Sako \etal 2002); they are
bound-bound transitions and their wavelengths and oscillator strengths
have been determined to very high accuracy for most elements (e.g.
Drake 1979, Savukov, Johnson, and Safronova 2002). 
However, there is very little reliable data in literature for 
resonances since they involve autoionizing 
states, not readily computed theoretically or observed experimentally.

 We refer to lines as due to transitions between initial and final bound 
levels, and to
resonances as due to transitions between a bound level and an
interacting bound and continuum state (quasi-bound). The resonance
may decay into the continuum with a finite lifetime, leading to a
resonance or autoionization width in energy; it may also decay radiatively
to a bound level as in the di-electronic recombination process.
In astrophysical observations, with resolution usually lower than typical
autoionization widths of the order of meV, line and resonance
features look similar. 
In X-ray work, resonances are often
referred to as `inner-shell lines', although resonances need not
be associated with inner-shells excitations alone. While the relevant
transition matrix elements may be computed by atomic structure codes by
explicitly coupling the bound and continuum wavefunctions in an isolated
resonance approximation, the
well established close-coupling approximation from
atomic collision theory enables a complete description of the coupled
wavefunctions and interference effects in resonances. Following
traditional atomic physics usage, we refer to a complex as a set of
lines or resonances belonging to the same principal quantum number.
(Eissner and Nussbaumer 1969, and references therein).

 Recently, theoretical calculations of wavelengths and resonance
oscillator strengths
for the O~VI KLL resonances (Pradhan 2000) predicted a feature at 22.05
$\AA$ that was later
detected for the first time in the \chan spectra of Seyfert 1 galaxy \mcg 
by Lee \etal (2001). 
Similar calculations have now been carried out for the Li-like ions, C~IV, 
O~VI, and Fe~XXIV (Nahar, Pradhan, and Zhang 2001). 
 These resonance oscillator strengths are needed in order to interpret 
the X-ray spectra of
many sources. Of prime interest is the plasma in
active galactic nuclei, photoionised by and surrounding the central
source. We may expect that {\it all} ionization states
of oxygen might be present in such sources, and should be 
identifiable through their \ka
resonance complexes, potentially leading to the determination of column 
densities and ionization fractions. Identifications of K- and L-shell
transitions in Oxygen and several other elements have also been made in the
\xmm spectra of Mrk 766 (Ogle \etal 2002, Mason \etal 2002), \mcg6 
(Branduardi-Raymont \etal 2001, Lee \etal 2001), and other sources 
(Kaastra \etal 2002, Blustin \etal 2002).
To enable the identification and analysis of such features, in
this brief {\it Letter} 
 we present the \ka resonance strengths derived from large-scale
relativistic close coupling calculations for the photoionization 
cross sections of all ions from O~I to O~VI.

\section{Theory and computations}

 A procedure for computing resonance oscillator strengths has been
described recently by Pradhan (2000).
 It is well known that we may relate the line oscillator strength and
the bound-free photoionization
cross section $\sigma_{PI}$ to the differential oscillator strength

\begin{equation}
\frac{df}{d\epsilon} = \left[ \begin{array}{l}
                         \frac{\nu^3}{2z^2} f_{line} \ \ \ \ ,
\hskip 2cm \epsilon < I \\
                         \frac{1}{4\pi^2 \alpha a_0^2} \sigma_{PI}
 \ , \hskip 2cm \epsilon > I
                         \end{array}
                         \right.
\end{equation}

 where I the ionization potential, z the ion charge,
$\nu$ the effective quantum number
at $\epsilon = -\frac{z^2}{\nu^2}$ in Rydbergs,
and  $\alpha$ and $a_0$ are the fine structure constant
and the Bohr radius respectively.
 The quantity $\frac{df}{d\epsilon}$ describes the strength of
photoabsorption
per unit energy, in the discrete
bound-bound region as well as the continuum bound-free region,
continuously across the ionization threshold. 

 The $\frac{df}{d\epsilon}$  reflects the
same resonance structure as the $\sigma_{PI}$ in the bound-free
continuum.
Combining the two forms of $\frac{df}{d\epsilon}$ we therefore define,
in the vicinity of a resonance, the integrated `resonance
oscillator strength' as:

\begin{equation}
 f_{r} (J_i \longrightarrow J_f)  =  \int_{\Delta E_{r}}
\left(
\frac{df (J_i \longrightarrow J_f)}{d\epsilon} \right) d\epsilon
= \left( \frac{1}{4\pi^2 \alpha
a_0^2} \right) \int \sigma_{\rm PI} (\epsilon; J_i \rightarrow J_f)
d\epsilon ,
\end{equation}

where $J_i,J_f$ represent the initial bound and the continuum symmetries.
 Eq. (2) may be evaluated from the detailed $\sigma_{PI}$ for the
symmetries concerned
provided the resonance profile is sufficiently well delineated. In
practice this is often difficult and elaborate methods need to be
employed to obtain accurate positions and profiles (the background and
the peaks) of resonances. Furthermore,
relativistic effects need to be included to differentiate the fine
structure components. Using the coupled channel formulation based
on the R-matrix and the relativistic Breit-Pauli R-matrix (BPRM) method
(Burke \etal 1971, Berrington \etal 1995) a large number of
photoionization cross
sections have been calculated for all astrophysically abundant elements
including resonance structures, particularly in the Opacity Project
and the Iron Project works (Seaton \etal 1994, Hummer \etal 1993).
 We compute the \dfe for all ions under
consideration from photoionization cross sections
from elaborate and extensive relativistic close coupling
calculations using the BPRM method.
In Eq. (2) the resonance oscillator strength \far is integrated
 over $\Delta E_r$, the energy range associated with the 
autoionization width(s). An advantage of the present method is that
\far may be computed not only for an isolated resonance, but also for
overlapping complex of resonances that may not be observationally resolved.

In the close-coupling or the coupled-channel approximation 
the bound or resonant \eion system is represented by a
wavefunction expansion
over coupled levels of the `core' or `target'
ion described by configuration-interaction type eigenfunctions. 
The presnet calculations for O-ions employ
expansions for initial ground states and 
\ka resonances associated with all different \eion continua (symmetry)
using a number of configuration-interaction type target ion configurations,
including those with 1s-hole in the K-shell and 2s-hole in the L-shell.
The extensive list of such spectroscopic configurations, and other  
computational details such as \eion short-range correlation
configurations, will be reported elsewhere. We note that these
close-coupling calculations are based on the theory of atomic
collisions, and are quite different from atomic structure
calculations generally employed to compute line oscillator strengths.

The resonant transitions from the ground state of an Oxygen
ion into the \ka autoionizing resonances are:
\be
\ba
 O~I: 1s^22s^22p^4   (^3P_2) + h\nu \lra 1s2s^22p^5 (^3P^o_{0,1,2},^1P^o_1) \\

 O~II: 1s^22s^22p^3  (^4S^o_{3/2}) + h\nu \lra 1s2s^22p^4 (^4P_{1/2,3/2/,5/2},
^2D_{3/2,5/2},^2S_{1/2},^2P_{1/2,3/2})\\

 O~III: 1s^22s^22p^2 (^3P_0) + h\nu \lra 1s2s^22p^3 (^5S^o_2,^3D^o_{1,2,3},
^1D^o_2,^3S^o_1,^3P^o_{0,1,2},^1P^o_1)\\

 O~IV: 1s^22s^22p (^2P^o_{1/2}) + h\nu \lra 1s2s^22p^2
(^4P_{1/2,3/2/,5/2},
^2D_{3/2,5/2},^2S_{1/2},^2P_{1/2,3/2})\\

 O~V: 1s^22s^2 (^1S_0) + h\nu \lra 1s2s^22p (^3P^o_{0,1,2},^1P^o_1) \\

 O~VI: 1s^22s (^2S_{1/2}) + h\nu \lra 1s2s2p (^2P^o_{1/2,3/2},^4P^o_{1/2,3/2,
5/2}). 

\ea
\ee

 The resonances on the RHS of (3) autoionise into a free
electron and a core ion in ground or excited states, consistent with
total fine structure \eion J$\pi$-symmetries given as subscripts of the
LS ($^{(2S+1)}L^{\pi}$) terms. 
 Whereas we locate the resonances due to all final continuum symmetries
on the RHS, some are overlapping or are weak with little integrated 
resonance oscillator strength. 
    
The monochromatic opacity due to the \ka resonances
may be obtained from the accurate bound-free photoionization
cross sections that delineate autoionization shapes or profiles.
Fine structure also needs to be included explicitly
since, as shown in this work, many fine structure components of
resonances may be present. Because several coupled continuum channels
contribute to each KLL resonance complex, as shown in Eq. (3), 
individual autoionization widths are not computed (and may be
overlapping); instead the whole
complex is delineated as function of energy. Integration over the entire
\ka photoabsorption range therefore yields the resonance oscillator strengths
reported in the next section.

Calculations for low ionization states 
are more difficult than for higher ones since the electron-electron
correlations are stronger relative to the dominant Coulomb potential,
and many more electronic configurations, both for the ion eigenfunctions
and for the \eion system described by continuum and bound channels,
need to be considered.

 Theoretical resonance spectroscopy, as formulated under the
present approach, yields precise
and detailed resonance positions, shapes, and oscillator strengths,
compelmenting traditional line spectroscopy.

\section{Results and discussion}

Fig. 1 shows the \ka resonant complexes of all O-ions O~I~-~OVI.
The dominant
components and the peak positions of resonances in each ion are shown.
As mentioned above, each complex has several components; their positions may or
may not overlap depending on the exact energies of the contributing
angular and spin symmetries. For example, O~I and O~II
show one single peak because the contributing final J$\pi$ symmetries
all lie
at the same energy ($\pi$ refers to the parity of the electronic state).
On the other hand resonances begin to spread out with ion charge
in O~III and O~IV, which also present more complicated resonance
structures. The ground level of O~III, $1s^22s^22p^2 (^3P_0)$,
photoionizes into the J = (1)$^o$ continuum of $1s2s^22p^3$ with the
three components shown. O~V has only one final J$\pi (^1P^o)$ since the
initial level is $1s^22s^2 \ (^1S_0)$. The KLL O~VI is a twin component
system discussed in detail in earlier works (Pradhan 2000, Nahar \etal
2001). However, the O~VI resonance at 21.87 \ang is about an order of
magnitude weaker than the stronger one at 22.05 \ang; the latter was
identified in the \mcg spectrum by Lee \etal (2001).
Most of these results for O-ions have been obtained for
the first time (prior works discussed below). 

 We note that the tentative identification by Lee \etal (2001) of the
O~I resonance in \mcg appears to be
slightly higher in energy than the present value, and
previous experimental and theoretical
values at $\sim 23.5$ \ang discussed by
Paerels \etal (2001), who used the non-relativistic
theoretical cross sections for O~I computed
by McLaughlin and Kirby (1998) to analyse interstellar X-ray absorption
(the theoretical model spectrum was shifted by
0.051 \ang to match the measured centroid wavelength of the O~I
resonance). Our O~II \ka resonance is about 0.08 \ang lower than
inferred from experimental data by Paerels \etal We also note that the computed
wavelength for O~V \ka resonance at 22.35 \ang is in good agreement with
the rest wavelength of 22.334 \ang recently observed with \xmm
by Blustin \etal (2002) from the Seyfert 1 galaxy NGC 3783.
O~III and O~IV resonances in Fig. 1 clearly illustrate the
overlapping and asymmetric nature of autoionizing resonance profiles of
the repective \ka complexes.

 In Table 1 we present the \ka resonance oscillator strengths for all
O-ions, most with \far $ > 0.1$. As evident from Fig. 1 
several components are nearly degenerate. The f$_r$ given in Table 1 are
therefore the sum over features lying approximately within $0.01 \AA$ of
each other. We also present autoionization `equivalent widths' W$_a$
(meV), together with maximum peak values $\sigma_{max}$, of the \ka 
resonances (note that \far = W$_a$ $\times \sigma_{max}$). 
It needs to be emphasized that while the values in Table 1
may be utilised to treat \ka resonances on par with `lines', a more
precise treatment is to consider the entire photoionization cross section,
up to energies encompassing the entire series of K-shell excitations, in
calculating the bound-free X-ray opacity. Such calculations are 
more extensive and are in progress.

 Another potentially important consideration is the level population in
excited fine structure levels of the ground LS term. If the source
plasma is moderately dense, $N_e > 10^6$ cm$^{-3}$, then the excited
levels may be significantly populated. Photoabsorption therefrom would
lead to additional \ka features to augment those presented herein.
These calculations are also in progress.

\section{Concluding remarks}

 The main conclusions of this work are as follows.

 1. The \ka resonance strengths of oxygen ions may be used to identify
and analyse soft X-ray spectral features around 0.5 - 0.6 keV, given the
\chan and \xmm resolution of $\sim 0.01 \AA$.

 2. Although not spatially co-existent, O~VI and O~VII regions may both 
be present and detectable via K-shell X-ray transitions from a given source.
 A Li-like ion, such as O~VI, 
is easily ionized, while the ionization of a He-like ion, such as O~VII, 
is the most energetic of all ionization states of an element and may
exist in a plasma over the widest temperature range (Pradhan 1982,1985). 
Thus the detection
of O~VI may be indicative of possibly a larger fraction of oxygen in
O~VII driven by recombination : (e~+~O~VII) $\lra$ O~VI.
The O~VI KLL doublet {\it absorption} resonances
at \lmlm  22.05 and 21.87 \ang (Pradhan 2000) lie
between the well known forbidden (`f'
or `z'), intercombination (`i' or `x,y'), and allowed (`r' or `w')
\ka {\it emission} lines of O~VII due to
transitions $2(^3S_1, ^3P^o_{2,1}, ^1P^o_1)
 \longrightarrow 1(^1S_0)$ at \lmlm 22.101, 21.804, and
21.602 \ang respectively (e.g. Ness \etal 2001).
To augment theoretical studies,
 new collisional calculations for electron impact excitation of O~VII
have been carried out (Delahaye and Pradhan 2002). In addition, self-consistent 
cross sections and rates O~VI/O~VII/O~VIII have been
computed (Nahar and Pradhan, in preparation) with the same
eigenfunction expansion for both photoionization and recombination, and
a unified (e~+~ion) recombination scheme including both radiative and
dielectronic recombination in an ab initio manner using the BPRM method
(e.g. Nahar and Pradhan 1992, Zhang \etal 1999).
The new O~VII total and level-specific recombination rates and
A-values, up to $n$ =
10 fine structure levels, should generally help the study of
recombination
lines, such as the O~VII `He$\beta, \gamma, \delta, \epsilon$'
identified by Sako \etal (2002) in the \xmm spectrum

 3. Analagous to the oxygen \ka resonances at $\sim$
0.55 keV (22-23 \ang), there are complexes corresponding
to other elements such as carbon at
$\sim$ 0.3 keV, nitrogen at $\sim$ 0.4 keV, neon at $\sim$ 0.9 keV. 
For example, at $\sim$ 0.94 keV Lee
\etal (2001) have identified a line from He-like Ne~IX ($1s^2 -
1s2p$). Calculations are in progress for these elements.

{\it Note to be added in proof:} A recent experimental and theoretical
study of \ka photoionization of O~II (Kawatsura \etal 2002) reports the
resonance oscillator strengths obtained from
multi-configuration Dirac-Fock (MCDF) calculations. Their combined value for 
the transitions $^4S_{3/2} - ^4P_{5/2}, ^4P_{3/2}, ^2P_{1/2}$ is 0.765,
compared to the present gf-value of 0.736 for the total \ka oscillator
strength. It is expected that the other transitions not considered by
Kawatsura \etal would be weaker, and if so then the present results should
agree with theirs to about a few percent. They also report that the 
measured position is 1.5
eV lower than the MCDF value; it appears that the energies in Table 1
and Fig. 1 for O~II may be higher by a similar amount, thus providing
an estimate of the uncertainties in the resonance positions in the
present work.

\section*{Acknowledgments}
This work was partially supported by the U.S. National 
Science Foundation and NASA Astrophysical Theory Program. 
The computational work was carried out on the Cray-SV1
at the Ohio Supercomputer Center in Columbus Ohio.

\def\amp{{ Adv. At. Molec. Phys.}\ }
\def\apj{{ApJ}\ }
\def\apjs{{ApJS}\ }
\def\aj{{AJ}\ }
\def\aa{{A\&A}\ }
\def\aasup{{A\&AS}\ }
\def\adndt{{ADNDT}\ }
\def\cpc{{Comput. Phys. Commun.}\ }
\def\jqsrt{{JQSRT}\ }
\def\jpb{{J. Phys. B}\ }
\def\pasp{{PASP}\ }
\def\mn{{MNRAS}\ }
\def\psc{{Phys. Scr.}\ }
\def\pra{{Phys. Rev. A}\ }
\def\prl{{Phys. Rev. Letts.}\ }

\newpage
\begin{table}
\caption{K$\alpha$ resonance oscillator strengths f$_r$ for oxygen ions}
\begin{center}
\begin{tabular} {lcccccc}
\hline
 Ion  & E$_r$ (Ryd) & E$_r$ (KeV) & $\lambda (\AA)$ & f$_r$ & W$_a$(meV)& 
$\sigma_{max}$ (MB) \\
\hline
 O I   & 38.8848  & 0.5288 & 23.45 & 0.113 & 31.88 & 48.05\\
\hline
 O II  & 39.1845  & 0.5329 & 23.27 & 0.184 & 23.21 & 107.7 \\

\hline
 O III & 39.5000  & 0.5372 & 23.08 & 0.119 & 27.00 & 59.92 \\
 O III & 39.6029  & 0.5386 & 23.02 & 0.102 & 12.08 & 114.5 \\
 O III & 39.7574  & 0.5407 & 22.93 & 0.067 & 24.27 & 37.48 \\ 

\hline
 O IV  & 40.1324  & 0.5458 & 22.73 & 0.132 & 27.11 & 66.15 \\ 
 O IV  & 40.2184  & 0.5470 & 22.67 & 0.252 & 14.32 & 239.2 \\
 O IV  & 40.5991  & 0.5521 & 22.46 & 0.027 & 21.91 & 17.00 \\ 

\hline
 O V   & 40.7826  & 0.5546 & 22.35 & 0.565 & 14.01 & 549.0 \\

\hline
 O VI  & 41.3456  & 0.5623 & 22.05 & 0.576 & 1.090 & 7142.0 \\
 O VI  & 41.6912  & 0.5670 & 21.87 & 0.061 & 12.16 & 67.36 \\

\hline 
\end{tabular}
\end{center}
\end{table}
\newpage
\begin{figure}
\centering
\psfig{figure=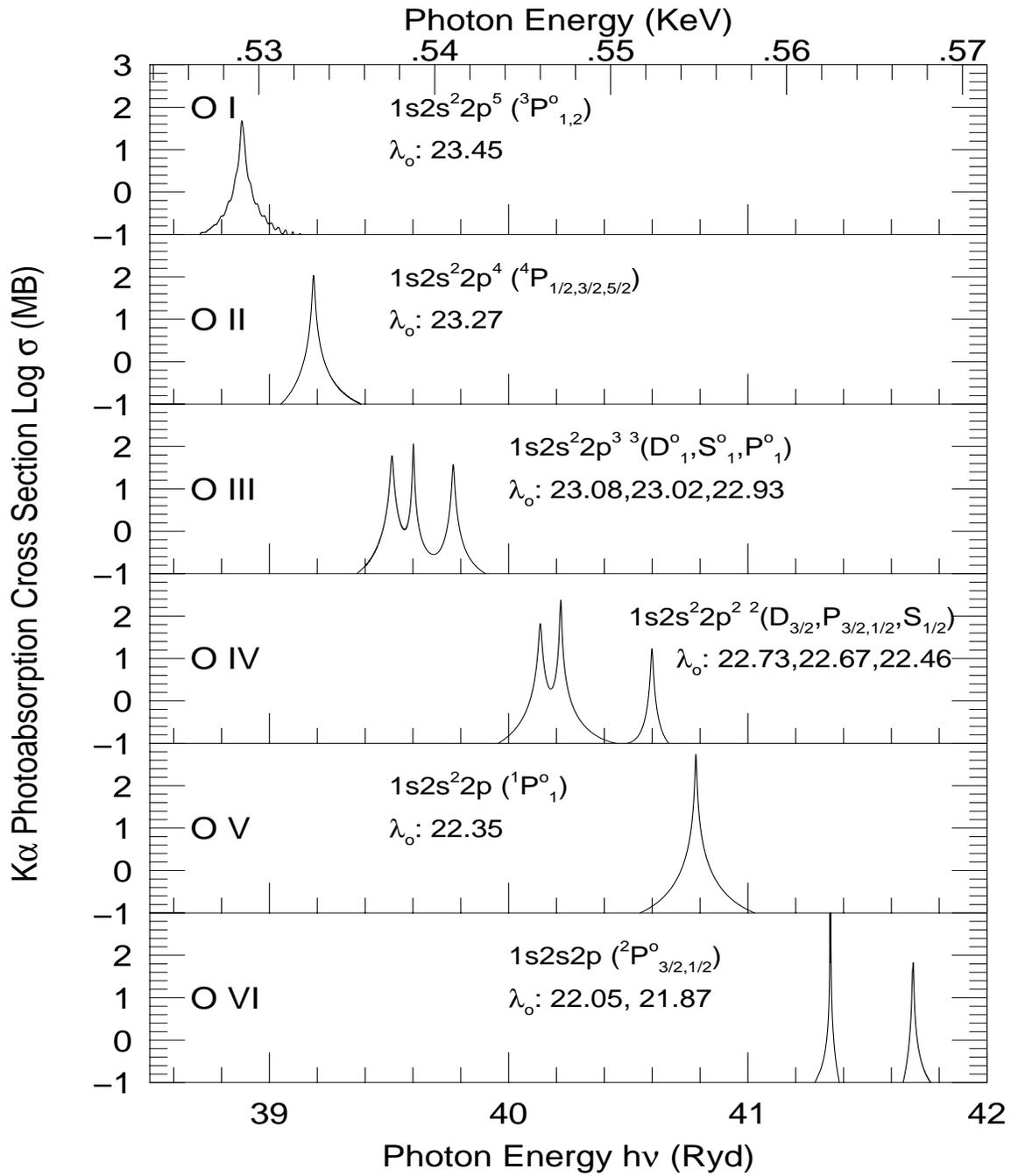,height=20.0cm,width=18.0cm}
\caption{Photoionization cross sections of \ka
resonance complexes in oxygen ions O~I~-~O~VI.}
\end{figure}

\end{document}